\documentstyle[12pt]{ioplppt}

\def\vec#1{{\bf#1}}
\def\hatn#1{\hat{\bf#1}}
\def\kf{k_{\mbox{\tiny{F}}}}
\def\pf{p_{\mbox{\tiny{F}}}}
\def\vf{v_{\mbox{\tiny{F}}}^*}
\def\mum{{\mu_{\mbox{\tiny M}}}}
\def\M2{\mbox{\tiny M$^2$RPA}}

\begin{document}

\jl{3}

\title{Towards a Fermi Liquid Theory of the $\nu=\frac{1}{2}$ State :
  Magnetized Composite Fermions}[Magnetized Composite Fermions]

\author{Steven H.  Simon\footnote{Work Done in Collaboration with Ady
    Stern and Bertrand Halperin}}

\address{Department of Physics, Massachusetts Institute of Technology,
  Cambridge, MA 02139}


\maketitle

\begin{abstract}
  The Fermionic Chern-Simons approach has had remarkable success in the
  description of quantum Hall states at even denominator filling fractions
  $\nu=\frac{1}{2m}$.  In this paper we review a number of recent works
  concerned with modeling this state as a Landau-Silin Fermi liquid.  We
  will then focus on one particular problem with constructing such a
  Landau theory that becomes apparent in the limit of high magnetic field,
  or equivalently the limit of small electron band mass $m_b$.  In this
  limit, the static response of electrons to a spatially varying magnetic
  field is largely determined by kinetic energy considerations.  We then
  remedy this problem by attaching an orbital magnetization to each
  fermion to separate the current into magnetization and transport
  contributions, associated with the cyclotron and guiding center motions
  respectively.  This leads us to a description of the $\nu=\frac{1}{2m}$
  state as a Fermi liquid of magnetized composite fermions which correctly
  predicts the $m_b$ dependence of the static and dynamic response in the
  limit $m_b \rightarrow 0$.  As an aside, we derive a sum rule for the
  Fermi liquid coefficients for the Chern-Simons Fermi liquid. This paper
  is intended to be readable by people who may not be completely familiar
  with this field.
\end{abstract}

\maketitle

\section{Introduction}

The Chern-Simons (or `composite') Fermion theory has had a number of
remarkable successes in the description of quantum Hall
states\cite{HLR,WillettReview}.  Based on the work of Jain\cite{Jain}, and
Zhang, Hansson, and Kivelson\cite{ZHK}, the Chern-Simons fermion picture
was first introduced by Lopez and Fradkin\cite{Lopez} to study
incompressible fractional quantized Hall states.  Later, in work by
Halperin, Lee and Read (HLR)\cite{HLR}, as well as Kalmeyer and
Zhang\cite{Zhang}, the theory was used to study even denominator filling
fractions.  A prediction of this approach is that the states at even
denominator filling fraction should be compressible Fermi liquid like
states.  However, several major problems have appeared in describing these
states as Fermi liquids.  Many of these problems are related to the
infra-red divergent properties of the Chern-Simons gauge field
fluctuations\cite{HLR,Ady,Kim,Millis,Kvesh}.  Recently, it has been
pointed out that there that there are also complications that are
unrelated to infra-red properties\cite{SimonStern}.  These complications
become most pronounced in the limit of large magnetic field (or
equivalently when the electron band mass $m_b$ is taken to zero).  A
resolution to the $m_b \rightarrow 0$ problems has been proposed in
Reference \cite{SimonStern} which involves binding of magnetization
(unrelated to spin) to each Chern-Simons quasiparticle.  The resulting
magnetized Fermi liquid description of even denominator Hall states yields
the correct behavior in the $m_b \rightarrow 0$ limit.

The current paper is written mainly to make the work of reference
\cite{SimonStern} more accessible to those who are not experts in the
field.  Thus, much background material will be discussed in detail.
In section \ref{sec:review} a brief review is given of previous works
relating to the Chern-Simons Fermionic picture of fractional Hall
states.  We begin by reminding the reader of a few essentials of
quantum Hall physics in section \ref{sec:Basics}.  In section
\ref{sec:CS} the basic Chern-Simons transformation is described in
detail and in section \ref{sec:MFT} the Chern-Simons mean field
description of both the incompressible fractional Hall states and the
compressible even denominator states is discussed.  Section
\ref{sec:Pert} is devoted to a brief review of several of the attempts
to perform a controlled perturbation theory around this mean field
solution.  We will also briefly mention some of the works that focus
on the infra-red divergences related to the gauge field fluctuations.

In section \ref{sec:Res} we define and discuss the electromagnetic
response functions $K$ and related response functions which are the
objects that we will attempt to calculate throughout the rest of the
paper.  The simplest and most commonly used approximation (beyond mean
field) for calculating these response functions is the Random Phase
Approximation (RPA).  This approximation will be discussed in section
\ref{sec:MRPA}.  It is pointed out that this approximation either
breaks Galilean invariance or incorrectly describes the energy scale
of the low energy excitations.  We then discuss how this problem is
corrected by using the Modified RPA (MRPA) from Reference
\cite{Simonhalp}.

In section \ref{sec:Mag} we discuss the physics of the large magnetic
field (or $m_b \rightarrow 0$) limit.  In particular, in section
\ref{sec:ZFR} we focus on the zero frequency, finite wavevector
electromagnetic response in this limit.  We show that the (M)RPA
incorrectly models some features of this response.  In section
\ref{sec:Binding} we propose that these problems can be repaired by
binding magnetization to each Chern-Simons quasiparticle.  In essence,
this binding allows for a separation of the current into a
magnetization current which is associated with the cyclotron motion of
electrons and a transport current associated with the guiding center
motion.  Following reference \cite{SimonStern}, in section
\ref{sec:M2RPA} a `Magnetized Modified RPA' ($\mbox{M}^2$RPA) is
defined that uses this magnetization binding approach in combination
with the MRPA to calculate the physical electromagnetic response
function $K$.

Section \ref{sec:FLT} is devoted to describing how this attachment of
magnetization (and the $\mbox{M}^2$RPA) fits into a Landau Fermi
liquid theory formalism.  In particular a new response function
($\tilde \Pi$) is defined that will give the self consistent response
for the magnetized quasiparticles.  Section \ref{sec:Boltzmann}
reviews Fermi liquid theory and defines the Boltzmann equation that
yields this response function as its solution.  In section
\ref{sec:sep} we separate out the effects of the Fermi liquid
coefficients that are singular in the limit $m_b \rightarrow 0$.  What
remains after this separation is then a Fermi liquid with reasonably
weak interactions.  In section \ref{sec:relation} we show that
approximating the response of this Fermi liquid as the response of
appropriate free fermions is precisely equivalent to the
$\mbox{M}^2$RPA.  Finally in section \ref{sec:conclusion} we make a
few additional comments and summarize our findings.  As an aside, in
Appendix A, a sum rule is derived for the Fermi liquid coefficients in
the Chern-Simons Fermi liquid.

\section{Review}
\label{sec:review}

\subsection{Basics}
\label{sec:Basics}

We begin by considering a system of $N$ interacting spin-polarized (or
spinless) electrons of band mass $m_b$ in a magnetic field $B = \nabla
\times \vec A$.  The Hamiltonian for this system is written as
\begin{equation}
  H = \sum_{j} \frac{\left[\vec p_j - \frac{e}{c} \vec A(\vec
    r_j)\right]^2}{2m_b} + \sum_{i<j} v(\vec r_i - \vec r_j).
\end{equation}
where $v$ is the two body interaction potential, $c$ is the speed of
light and $e$ is the charge of the electron.  We will often specialize
to the physical case of Coulombic interaction $v(r) = e^2/(\epsilon
r)$ with $\epsilon$ the background dielectric function.  However, it
will also be useful at times to consider other forms of
electron-electron interaction.

Ignoring interactions between the electrons, the single particle
spectrum breaks up into Landau-levels with energy $E_n = \hbar
\omega_c (n + \frac{1}{2})$ where
\begin{equation}
  \omega_c = \frac{eB}{m_b c}
\end{equation}
is the cyclotron frequency.  Each such Landau band has a degeneracy of
$B/\phi_0$ per unit area where
\begin{equation}
  \phi_0 = \frac{2 \pi \hbar c}{e}
\end{equation}
is the flux quantum.  The filling fraction
\begin{equation}
  \label{eq:nudef}
  \nu = \frac{\phi_0 n_e}{B}
\end{equation}
where is the $n_e$ the electron density thus gives the number of
Landau levels completely filled.  Note that when an integer number of
Landau bands are completely filled (ie, $\nu$ is an integer), there is
a discontinuity in the chemical potential leading to an incompressible
integer quantized Hall state\cite{Prange}.

When $\nu$ is a fraction (particularly for $\nu < 1$), due to the
degeneracy of single particle states, the physics is controlled by the
inter-electron interaction.  We note that the interaction energy scale
is given by $v(l_0)$ where $l_0 = \sqrt{\phi_0/(2 \pi B)}$ is the magnetic
length.  In the large magnetic field limit (or equivalently when $m_b
\rightarrow 0$), the interaction energy scale is much less than the
cyclotron scale. However, due to the large degeneracy of states,
traditional perturbation methods in the interaction $v$ are not
effective for $\nu < 1$.  In order to understand this regime, we will
use the Chern-Simons transformation described below.

\subsection{Chern-Simons Transformation}
\label{sec:CS}

Writing the electron wavefunction $\Phi(z_1,z_2,\ldots,z_N)$ with $z_j
= x_j + i y_j$ the position of the $j^{th}$ electron, it can be shown
that\cite{Lopez,HLR} if $\Phi$ is a solution of the Schroedinger
equation $H \Phi = E \Phi$, then for $m$ an integer,
\begin{equation}
  \label{eq:noanal1}
  \Psi(z_1,z_2,\ldots,z_N) = \prod_{i<j} \left[ \frac{(z_i - z_j)}{|z_i -
    z_j|} \right]^{2m} \Phi(z_1,z_2,\ldots,z_N)
\end{equation}
is a solution to the Schroedinger equation $H' \Psi = E \Psi$ with
\begin{equation}
  H' = \sum_{j} \frac{\left[\vec p_j - \frac{e}{c} \vec A(\vec r_j) +
    \frac{e}{c} \vec a(\vec r_j)\right]^2}{2m_b} + \sum_{i<j} v(\vec
  r_i - \vec r_j)
\end{equation}
the Hamiltonian for $N$ interacting fermions where $\vec a$ is the
`Chern-Simons' vector potential
\begin{equation}
\label{eq:csvec1}
\vec a(\vec r) = \frac{\tilde \phi \phi_0}{2 \pi} \sum_{j=1}^{N}
\frac{\hatn{z} \times (\vec r - \vec r_j)}{|\vec r - \vec r_j|^2},
\end{equation}
and $\tilde \phi = 2m$.  The Chern-Simons magnetic field $b(\vec r)$
associated with the vector potential $\vec a$ is given by
\begin{equation}
\label{eq:last1}
b(\vec r) = \nabla \times \vec a(\vec r) = \tilde \phi \phi_0
\sum_{j=1}^N \delta(\vec r - \vec r_j) = n(\vec r) \tilde \phi \phi_0.
\end{equation}
where $n(\vec r)$ is the local particle density.  In other words, the
Chern-Simons transformation can be described as the exact modeling of
an electron as a fermion attached to $\tilde \phi = 2m$ flux quanta.
We call these fermions `gauge transformed', `composite', or
`Chern-Simons' fermions\footnote{Note that term `composite fermion' is
  used by Jain\cite{Jain} in a somewhat different sense}.

\subsection{Mean Field Theory}
\label{sec:MFT}

The simplest approach to analyzing this system is to make the mean
field approximation in which density is assumed uniform and the
Chern-Simons flux quanta attached to the fermions are smeared out into
a uniform magnetic field of magnitude
\begin{equation}
  \label{eq:mfield1}
  \langle b \rangle = n_e \tilde \phi \phi_0
\end{equation}
with $n_e$ the average density, and $\tilde \phi = 2m$ again. Choosing
the Chern-Simons flux to be in the opposite direction as the applied
magnetic field, at some special value of the filling fraction, when $B
= \langle b \rangle$, the applied magnetic field precisely cancels the
Chern-Simons flux at the mean field level.  This exact cancelation
occurs at the filling fraction
\begin{equation}
\nu = \frac{n_e \phi_0}{ \langle b
  \rangle} = \frac{1}{2m}
\end{equation}
At these special filling fractions, the mean field system can be
described as fermions in zero magnetic field, and should therefore be
a compressible Fermi-liquid like state.  The existence of this
Fermi-liquid like state at even denominator filling fractions was
predicted by Kalmeyer and Zhang\cite{Zhang} and by Halperin, Lee, and
Read\cite{HLR}.  It should be noted that this mean field description
of the $\nu=\frac{1}{2m}$ state is a nondegenerate starting point for
attempting a controlled perturbation theory --- unlike the original
highly degenerate Landau Levels.

For completeness, we also consider the case when the filling fraction
is away from $\nu = \frac{1}{2m}$.  Here, the applied magnetic field
and the Chern-Simons flux do not cancel.  At the mean field level, a
residual field
\begin{equation}
  \label{eq:DeltaB1}
  \Delta B = B - \langle b \rangle = B - \tilde \phi n_e \phi_0 = B -
  2 m n_e \phi_0
\end{equation}
is left over.  Thus, the mean field system is described as
noninteracting fermions in the uniform field $\Delta B$.  The
effective filling fraction for these gauge transformed fermions is
given by
\begin{equation}
\label{eq:onepoint}
p = \frac{n_e \phi_0}{\Delta B.}
\end{equation}
When $p$ is a small integer, at the mean field level, this is just a
system of $|p|$ filled Landau levels of fermions, and one should
observe the integer quantized Hall effect of transformed fermions.
Using Eq.  \ref{eq:DeltaB1} as well as the definition of the filling
fraction (Eq. \ref{eq:nudef}), this condition (\ref{eq:onepoint})
yields precisely the Jain series\cite{Jain} of fractional quantized
Hall states
\begin{equation}
  \nu = \frac{p}{2mp+1.}
\end{equation}
Thus, the fractional quantized Hall effect at these filling fractions
is identified with an integer quantized Hall effect of gauge
transformed fermions\cite{Lopez}.  The excitation gaps for these
quantized Hall states are naturally given by the corresponding
effective cyclotron frequency of the composite fermions
\begin{equation}
  \label{eq:gap}
  E_g = \hbar \Delta \omega_c^* = \frac{\hbar e \Delta
    B}{m_{gap}^*(\nu) c}
\end{equation}
where $m_{gap}^*(\nu)$ is an effective mass to be discussed further
below.

\subsection{Perturbative Approaches}
\label{sec:Pert}

Although at a mean field level, the $\nu = \frac{1}{2m}$ system looks
like a Fermi liquid, we do not expect such a simple mean field
approximation to accurately describe the system.  Previous attempts
for going beyond mean field theory have so far involved perturbative
treatments of the Chern--Simons and electrostatic
interactions\cite{HLR,Ady,Kim,Millis,Kvesh,Simonhalp}.  There are
several major difficulties in these approaches.  To begin with, the
`small' dimensionless parameter that one must use in the perturbation
theory is $\tilde \phi = 2m \ge 2$ which is by no means small.  So
although the mean field solution seems like a good starting point for
a controlled perturbation theory, the remaining interactions are quite
strong and are not in the perturbative regime.

Furthermore, even if $\tilde \phi$ were small there would still be
problems with the perturbative treatment of the Chern-Simons
theory\footnote{Perturbing in $\tilde \phi$ can be considered
  appropriate for the modeling of a system of {\it anyons} with
  statistical angle $\theta$ in a magnetic field $B = \theta n
  \phi_0/(2 \pi)$ (Here fermions are defined to have statistical angle
  0 modulo $2 \pi$).  By similarly attaching $\tilde \phi =
  \theta/\pi$ quanta of flux to each particle, we obtain a system that
  in mean field theory is described as Fermions in zero field.  This
  family of anyonic systems with different $\theta$ parameters
  presumably share many similar properties.  So long as no phase
  transitions occur between $\theta =0$ and $\theta = 2 \pi$, the
  properties of the composite fermion system ($\theta = 2 \pi$ or
  $\tilde \phi = 2$) should be qualitatively described by perturbation
  theory in $\tilde \phi$}.  One problem that has attracted much
attention arises when the electrostatic interaction $v(r)$ is of
Coulomb form or is shorter ranged.  If this is the case, it is found
that composite fermion's effective mass at the Fermi surface diverges,
due to infra-red gauge field fluctuations\cite{HLR,Ady,Kim}.  Although
this divergence is reflected in the energy gaps (See Eq. \ref{eq:gap})
of fractional quantized Hall states at $\nu=\frac{p}{2mp+1}$ (for
large $p$)\cite{Ady}, the diverging effective mass is thought not to
affect the electronic linear response at $\nu=\frac{1}{2m}$ at zero
temperature, due to a mutual cancelation with another singular
term\cite{Ady,Kim,Millis,Kvesh}.  Consequently\cite{Ady} the low
energy excitations at $\nu = \frac{1}{2m}$ are best characterized by
another, {\it finite}, effective mass, denoted by $m^*$, which is the
effective mass of relevance to the present work.  It is this $m^*$
which should determine the scale of the fractional Hall gaps for small
values of $p$.

In order to avoid the complications associated with this divergence,
we can consider in this paper a system with interactions that are
longer ranged than Coulomb such that there are no infra-red
divergences.  (The long range interaction suppresses density
fluctuations and hence kills the effects of the gauge field at long
distances).  However, due to the above mentioned cancelation of
divergences in physical response functions\cite{Ady,Kim,Millis,Kvesh},
we believe that the conclusions reached below will be independent of
the range of the interaction.

In section \ref{sec:Mag} and \ref{sec:FLT} below we will address a
completely independent problem that occurs in the limit of $m_b
\rightarrow 0$ (or equivalently for large magnetic field $B$).  In
this limit the ground state and low energy excitations are
constrained to the lowest Landau level.  This lead to restrictions on
the electromagnetic response that are not properly described by simple
perturbative approaches.  In section \ref{sec:Res} below, we will
define this response function, and in section \ref{sec:MRPA} we will
describe the simplest approaches for going beyond mean field --- the
RPA and MRPA approximations.  Finally, in section \ref{sec:Mag} we
will show why these approximations are are insufficient in the $m_b
\rightarrow 0$ limit.

\subsection{Response Functions}
\label{sec:Res}

The quantity that we will attempt to calculate is the electromagnetic
response matrix $K_{\mu\nu}$ which is closely related to the
conductivity\cite{HLR,Simonhalp} (See Eqns. \ref{eq:KPi},
\ref{eq:Pirho}, and \ref{eq:rhorho} below).  To define $K$, a weak
vector potential $A_\mu^{\mbox{\scriptsize{ext}}}$ is externally
applied to a system at wavevector ${\bf q}$ and frequency $\omega$,
and consequently, a current $j_\mu$ is induced (Here $A_0$ is the
scalar potential, and $j_0$ is the induced density).  We write the
response function in the form
\begin{equation}
  \label{eq:Kdef}
j_{\mu}(q, \omega) = K_{\mu \nu}(q, \omega)
A_{\nu}^{\mbox{\scriptsize{ext}}}(q, \omega)
\end{equation}
where $\mu$ and $\nu$ take the values $0,x,y$.  We will use the
convention that the perturbation is applied with $q \| {\bf {\hat x}}$
so that the longitudinal current is $j_x = (\omega/q) j_0$.  Using the
gauge ${\bf A}_x = 0$, we can then treat $K_{\mu\nu}$ as a $2\times 2$
matrix with indices taking the values 0 or 1 denoting the time or
transverse space components.  In this notation the current vector
$j_\mu$ is $(j_0,j_y)$, and the vector potential $A_\mu$ is
$(A_0,A_y)$.  Note that from here on, we will routinely drop the
explicit matrix subscripts $\mu$ and $\nu$ as well as the explicit $q$
and $\omega$ dependences.

In systems with long ranged Coulomb interactions, a density $j_0(\vec
q)$ induced by the external vector potential, gives rise to an
additional Coulomb scalar potential $ev(q)j_0(\vec q)$, where $v(q) =
\frac{2\pi}{\epsilon q}$ is the Fourier transform of the usual Coulomb
interaction $v(r) = \frac{1}{\epsilon r}$ (with $\epsilon$ the
background dielectric constant). Similarly, for the Chern-Simons
fermion theory of the $\nu = \frac{1}{2m}$ state, in addition, an
induced vector potential originates from the composite fermions' flux.
An excess density $j_0(\vec q)$ carries an excess flux
$2\pi\tilde{\phi} j_0(\vec q)$ with $\tilde \phi = 2m$. A composite
fermions' current $\vec j(\vec q)$ is a current of flux tubes,
inducing an electric field $2\pi\tilde{\phi} \vec j(\vec q)$. Thus,
the composite fermions' current induces also a vector potential.
Keeping a matrix notation, we may write the induced vector potential
as
\begin{equation}
  \label{eq:AUJ}
  A^{\mbox{\scriptsize{ind}}} = U j
\end{equation}
where
\begin{equation}
\label{eq:Udef}
U=\left[ \begin{array}{cc} v(q) & 0 \\ 0 & 0 \end{array} \right] +
\frac{2 \pi \tilde \phi \hbar}{e} \left[ \begin{array}{cc} 0 &
-\frac{i}{q}
  \\ \frac{i}{q} & 0 \end{array} \right]
\end{equation}
where the first term is the Coulomb contribution and the second term
is the Chern-Simons contribution.  (We have now dropped the explicit
$q$ and $\omega$ dependences as well as the matrix subscripts in Eq.
\ref{eq:AUJ}.

Above, we have discussed the electromagnetic response function $K$
which gives the current response to the externally applied vector
potential.  It is now useful to define another response function
$\Pi$, which relates the current $j_\mu$ to the {\it total} vector
potential\footnote{Our matrix $\Pi$ is written as $\tilde K$ in
  references \cite{HLR} and \cite{Simonhalp}.  However, our notation
  for $\Pi$ agrees with that used in references \cite{Ady}, \cite{Kim}
  and \cite{SimonStern}},
\begin{equation}
  \label{eq:Pidef}
  j = \Pi A^{\mbox{\scriptsize{total}}}
\end{equation}
with
\begin{equation}
  \label{eq:Atotal}
  A^{\mbox{\scriptsize{total}}} = A^{\mbox{\scriptsize{ext}}} +
  A^{\mbox{\scriptsize{ind}}}
\end{equation}
so that
\begin{equation}
  \label{eq:KPi}
  K^{-1} = \Pi^{-1} + U
\end{equation}
Thus $\Pi$ is the part of $K$ that is irreducible with respect to both
Coulomb and Chern--Simons interactions.

The matrix $\Pi$ also defines the finite frequency and wavevector
composite fermion resistivity\footnote{In references \cite{HLR} and
  \cite{Simonhalp} $\rho_{cf}$ is called $\tilde \rho$} $\rho_{cf}$
via
\begin{equation}
\label{eq:Pirho}
 \rho_{cf}  =  [ T \Pi T ]^{-1}
\end{equation}
where $T$ is the conversion matrix
\begin{equation}
  \label{eq:TT}
  T= \left[ \begin{array}{cc} \frac{i\sqrt{i\omega}}{q} & 0 \\ 0 &
  \frac{1}{\sqrt{i\omega}}
\end{array} \right]_.
\end{equation}
The composite fermion resistivity $\rho_{cf}$ is the matrix that
relates the $\hatn x$ and $\hatn y$ components of the total (induced
and external) electric field $\vec E^{\mbox{\scriptsize{total}}}$ to
the $\hatn x$ and $\hatn y$ components of the current $\vec j$ via the
$2 \times 2$ matrix equation
\begin{equation}
  \label{eq:ejcf}
  \vec E^{\mbox{\scriptsize{total}}} = \rho_{cf} \vec j.
\end{equation}
where $\vec E^{\mbox{\scriptsize{total}}}$ is the electric field
associated with the vector potential $A^{\mbox{\scriptsize{total}}}$.
Equation \ref{eq:Pirho} simply converts $\Pi$ to $\rho_{cf}$ by using
appropriate factors of $\omega$ and $q$ to convert $\vec
E^{\mbox{\scriptsize{total}}} $ to $A^{\mbox{\scriptsize{total}}}$,
and $j_0$ to $j_x$.

In terms of this composite fermion resistivity, the original electron
resistivity $\rho$ (at finite $q$ and $\omega$) is given
by\cite{KLZ,Zhang,HLR}
\begin{equation}
  \label{eq:rhorho}
  \rho = \rho_{cf} + \rho_{CS}
\end{equation}
with
\begin{equation}
  \rho_{CS} = \frac{2 \pi \hbar \tilde \phi}{e^2} \left[
  \begin{array}{cc} 0 & 1 \\ -1 &
    0
\end{array} \right]
\end{equation}

\subsection{RPA and Modified RPA (MRPA)}
\label{sec:MRPA}

In order to find the electromagnetic response $K$ at even a crude
level, we must account for the interactions (both Coulomb and
Chern-Simons) beyond mean field.  The simplest approach to account for
these interactions is the Random Phase Approximation (RPA).  Making
the separation of $K$ into $\Pi$ and $U$ as described above in Eq.
\ref{eq:KPi}, the RPA approximation is equivalent to approximating
$\Pi$ as the response $K^0$ of noninteracting electrons of mass $m_b$
in the (mean) uniform magnetic field $\Delta B$.  Such an
approximation was originally discussed by Lopez and Fradkin for the
Jain series of fractional quantized Hall states\cite{Lopez} and by
Halperin, Lee, and Read\cite{HLR} and Kalmeyer and Zhang\cite{Zhang}
for the even denominator states.  In terms of resistivities, the RPA
amounts to defining the composite fermion resistivity $\rho_{cf}$ to
be the resistivity for a system of free fermions with mass $m_b$.

As pointed out in reference \cite{HLR}, if one makes this RPA
approximation and in the calculation of $K^0$ one uses the bare band mass
$m_b$, then, at least at mean field level, it is this mass that determines
the scale of the low energy excitations (ie, $m^*_{gap}(\nu) = m_b$ in Eq.
\ref{eq:gap}).  Since the low energy excitations should be controlled by
the interaction strength, this is clearly incorrect.  Of course, if one
could properly treat the fluctuations of the gauge field, presumably the
scale of the low energy excitations would indeed be found to be on the
interaction scale\footnote{Note that in the Chern-Simons boson model of
  the fractional quantized Hall effect, properly treating the vortex
  configurations of the superfluid can be shown to give the low energy
  excitations correctly on the interaction scale\cite{ZhangReview}}.  We
note, however, that at the present no approximation is known that properly
achieves the low energy excitaton scale by including fluctuations.  Thus,
a realistic approximation must have this low energy excitation scale
repaired by hand.

The simplest way to repair the problem of having low energy excitations on
the wrong energy scale is to phenomonologically approximate $\Pi$ as
$K^{0*}$, the response of a system of noninteracting electrons in the mean
magnetic field $\Delta B$ with a new effective mass $m^*$, where $m^*$ is
some phenomenological effective mass set by the interaction
scale\cite{HLR} (so $1/m^* \sim e^2/(\epsilon l_0)$).  For typical
experimental parameters, the measured effective mass is on the order of 4
to 15 times that of the bare band mass\cite{HLR,WillettReview}.
Unfortunately, simply replacing $m_b$ by $m^*$ leads to a theory with
several serious problems.  The strategy we will generally employ is to
adopt this mass replacement, identify the resulting problems and find ways
to repair them phenomonologically.  Once again we note that if we had a
way to properly treat the gauge field fluctuations such that the low
energy excitations were naturally on the interaction scale, we would not
have the problems that we will discuss and attempt to repair below.

To begin with, it can be shown that the naive replacement of $m_b$ by
$m^*$ results in a theory that violates Galilean
invariance\cite{Simonhalp}.  In particular, Kohn's theorem (a result of
Galilean invariance) requires that the only excitation mode with weight in
the long wavelength limit is the cyclotron mode at frequency $\omega_c = e
B/m_b c$.  This mode is a reflection of the oscillation of the center of
mass of the entire system and must therefore be independent of
interactions.  If one naively replaces $m_b$ by $m^*$, once ends up with a
cyclotron mode instead at the incorrect renormalized cyclotron frequency
$e B/m^* c$.  Similarly, simply replacing $m_b$ by $m^*$ results in a
violation of the so-called $f$-sum rule\cite{Simonhalp}.  We will show
later in section \ref{sec:Mag} that this replacement of the band mass with
the effective mass has a number of additional effects that need to be
properly treated before we obtain a fully viable phenomological theory.

In Reference \cite{Simonhalp} a Modified RPA (MRPA) was constructed
that restores Galilean invariance while keeping the low energy
excitations on the interaction scale.  In this MRPA approximation, the
mass renormalization from $m_b$ to $m^*$ is compensated for by
including a Fermi liquid interaction coefficient ${\cal F}_1$ (this
will be discussed further below).  To define the MRPA, we write,
\begin{eqnarray}
  \label{eq:Pi*}
  \Pi^{-1} &=& [\Pi^*]^{-1} + {\cal F}_1 \\
  \label{eq:F1def} {\cal F}_1 &=&
  {\frac{(m^*\!-\!m_b)}{n_e e^2}} \left(
  \begin{array}{cc}  {\frac{\omega^2}{q^2}} & 0 \\ 0 & -1 \end{array}
  \right)_.
\end{eqnarray}
The MRPA is then obtained by setting $\Pi^*$ equal to the response
$K^{0*}$ of a system of noninteracting fermions of mass $m^*$ in the
mean magnetic field $\Delta B$.  The response function thus calculated
(using $\Pi^* =K^{0*}$ and Eqns.~\ref{eq:Pi*} and \ref{eq:KPi}) will
be called $K^{{\mbox{\tiny{MRPA}}}}$.  Note that the form of Eq.
\ref{eq:Pi*} is similar to that of Eq. \ref{eq:KPi} in the sense that
it separates out the effect of an interaction term.  Similar to the
RPA approach of Eq. \ref{eq:KPi}, here ${\cal F}_1$ is an effective
interaction and $\Pi^*$ is a response function calculated without the
interaction ${\cal F}_1$ included.  Comparisons of results of exact
diagonalizations of small systems projected to the Lowest landau level
to results of $K_{00}$ calculated in the MRPA were quite
favorable\cite{SH} for the low energy excitations at
$\nu=\frac{p}{2mp+1}$ for small $p$.  Similar comparisons at
$\nu=\frac{1}{2}$ also yielded favorable results for small
systems\cite{Morf}.  Despite these successes, we will show below that
the (M)RPA does not properly represent the other elements of the
response matrix ($K_{01},K_{10}$, and $K_{11}$) in the limit of $m_b
\rightarrow 0$.

\section{Magnetized Fermions}
\label{sec:Mag}

We now turn to consider the limit of small band mass $m_b$ (or
equivalently large magnetic field $B$).  The fact that, in this limit, the
electronic ground state and low energy excitations are constrained to the
lowest Landau level, leads to certain features of the electronic response
to an external static vector potential which are not properly represented
in approximation schemes such as the mean field or the (M)RPA if we have
used a renormalized mass $m^*$ to achieve the correct energy scale for low
energy excitations.  We note that this problem occurs in the Chern-Simons
theory even when gauge-field fluctuations are not infra-red singular.
(For example, if the electron-electron repulsion falls off more slowly
than $1/r$ there should be no infra-red divergences in the effective
mass).

In reference \cite{SimonStern}, a new approach is proposed that is
based on a separation of the current into a magnetization current
which is associated with the cyclotron motion of electrons and a
transport current associated with the guiding center motion.  This
separation is achieved by attaching a magnetization ${\mu_{\mbox{\tiny
      M}}}$ to each particle.  This magnetization originates from the
electrons' orbital motion and is unrelated to the spin (we have
assumed spinless electrons throughout this paper).  In the limit $m_b
\rightarrow 0$, the magnetization ${\mu_{\mbox{\tiny M}}}$ is given by
the Bohr magneton
\begin{equation}
  \mu_b = \frac{e \hbar}{2 m_b c}
\end{equation}
The proposed separation procedure combined with approximations similar
to those made in the MRPA, results in an approximation we call the
$\mbox{M}^2$RPA that yields a response functions that correctly
describes the $m_b \rightarrow 0$ limit.

\subsection{Zero Frequency Response}
\label{sec:ZFR}

In this section we shall examine the form of the zero frequency finite
wavevector response in the high magnetic field (or $m_b \rightarrow
0$) limit.  An acceptable approximation for calculating the response
of the $\nu= \frac{1}{2m}$ state must correctly predict this limit.
We will show below that the usual Chern-Simons approaches do not
correctly predict this limit.  We then discuss in section
\ref{sec:M2RPA} below how the magnetization attachment proposed in
reference \cite{SimonStern} corrects this problem.

Consider the $\nu=\frac{1}{2m}$ state in the limit $m_b \rightarrow
0$.  In this limit the gap between Landau levels becomes large so we
expect such a system to be restricted to the lowest Landau level.  If
we apply a weak external static scalar potential at wavevector $q$ to
the system, the resulting state should remain in the lowest Landau
level so the induced density fluctuation should depend only on the
interaction strength, and not on the bare mass $m_b$.  Thus, $K_{00}$,
the so-called density-density response, should be independent of the
bare mass in this limit (or more properly, should scale as $(m_b)^0$
plus ${\cal O}(m_b)$ corrections).  However, the resulting density
inhomogeneity will yield a transverse current called the
magnetization current, given by  (here and below the speed of light
$c=1$)
\begin{equation}
  \label{eq:magM}
  \vec j_{\mbox{\scriptsize{mag}}} = \hatn z \times \nabla \vec M
\end{equation}
with $\vec M$ the magnetization density.  For noninteracting particles
in the lowest Landau level, the kinetic energy density is
\begin{equation}
  E \equiv \vec M \cdot \vec B = \frac{1}{2} \hbar \omega_c n_e
\end{equation}
so that the magnetization per particle is $|\vec M|/n_e= \mu_b$, the Bohr
magneton.  More generally, when interactions are taken into account, we
let the magnetization per particle be given by a quantity $\mum$ which
must become $\mu_b$ in the $m_b \rightarrow 0$ limit where the system
becomes projected to the Lowest Landau level.  We can thus
write\cite{Girvin} the magnetization current as\footnote{When projected to
  the lowest Landau level, the projected current and density operators
  satisfy $P {\bf j} P = \mu_b ({\bf {\hat z}} \times \nabla P n P)$ where
  $P$ is the projection operator.  In other words, for projected states,
  all of the current is magnetization current.}
\begin{equation}
  \label{eq:magn}
  \vec j_{\mbox{\scriptsize{mag}}} = \mum (\hatn z \times \nabla n)
\end{equation}
with $n(\vec r)$ the local electron density.  The physical
interpretation of this magnetization current as follows.  Each
particle in the lowest Landau level can be thought of as a particle in
a cyclotron orbit.  When the density of particles is uniform, the
local currents of all of these orbits cancel and there is no net
current in the system.  However, when there is a density
inhomogeneity, these local currents do not quite cancel and a net
magnetization current results.  Note that this magnetization current
associated with density gradients can be modeled by imagining that a
small magnetization $\mum$ (equivalent to a current loop) is attached
to each quasiparticle.

Using Eq. \ref{eq:magn} we see that in the limit $m_b \rightarrow 0$,
when we apply the weak static scalar potential
$A_{0}^{\mbox{\scriptsize{ext}}}(q)$ to the system and we look at the
leading current response we find a magnetization current $\mu_b
{\bf{\hat z}} \times i {\bf q} K_{00}
A_{0}^{\mbox{\scriptsize{ext}}}$.  Thus, if $q$ is finite we expect
\begin{equation}
\label{eq:Klim}
\lim_{m_b \rightarrow 0} K_{10}/K_{00} = i q \mu_b
\end{equation}
This result is not contained in works based on the Chern-Simons
approach previous to that of Reference \cite{SimonStern}.

We can also consider applying a weak external transverse vector
potential $A_1^{\mbox{\scriptsize{ext}}}$ at wavevector $q$ and zero
frequency.  This transverse field generates a magnetic field $\delta B
= i qA_1$ at wavevector $q$.  The variation in the total magnetic
field $B(\vec r) = B_{1/2} + \delta B(\vec r)$ will make the kinetic
energy $\frac{1}{2}\hbar \omega_c(\vec r) = \mu_b B(\vec r)$
positionally dependent thus attracting electrons to the regions of
minimal magnetic field when $m_b \rightarrow 0$.  This attraction is
not modeled in the Chern-Simons fermion picture at the mean field or
(M)RPA level if a renormalized mass is used.

Formally, if the applied variation in magnetic field generates a
density fluctuation $j_0(q)$, we can write the energy cost as
\begin{equation}
  \label{eq:Ener1}
  \delta E = j_0 (\delta B) \mum + \frac{1}{2} K_{00} j_0^2
\end{equation}
where $K_{00}$ is independent of $m_b$ as discussed above.  The first
term here is just the change in local cyclotron energy which can be
thought of as an effective scalar potential for the fermions.  This
term would occur quite naturally if we were to imagine that a
magnetization $\mum$ were attached to each fermion.  The second term
in Eq.  \ref{eq:Ener1} is due to the Coulomb interactions within the
lowest Landau level.  Again note that $\mum$ must become $\mu_b$ in
the $m_b \rightarrow 0$ limit, but more generally can include pieces
on the interaction scale.

Minimizing the energy (Eq. \ref{eq:Ener1}) with respect to $j_0$
yields the density
\begin{equation}
  j_0 = - (\delta B) \mu_m K_{00} = - i q \mum K_{00} A_1
\end{equation}
from which we conclude that that the leading term of $K_{01}$ is given
by $i q \mum K_{00}$ (in accordance with the symmetry requirement of
the matrix $K$).

Finally, once we have determined the density fluctuation due to this
local magnetic field fluctuation, we again realize that this density
fluctuation results in a magnetization current, so that we have a
leading piece of $K_{11}$ given by $K_{00} q^2 \mum^2$.

\subsection{Binding Magnetization to Composite Fermions}
\label{sec:Binding}

As suggested by the above discussion, the necessary correction to the
composite fermion picture involves attaching a magnetization $\mum$ to
each composite fermion so that it properly represents a particle in
the lowest Landau level.  Attaching magnetization to each particle can
also be interpreted as attaching a current loop to each particle
associated with the electrons' cyclotron motion. Thus the total
current would include both a piece from the motion of the
particle-currentloop composite and a piece from the current loop
itself. To this end, we define a transport current\footnote{The
  division into ${\bf j}_{\mbox{\tiny{trans}}}$ and ${\bf
    j}_{\mbox{\tiny{mag}}}$ has some degree of arbitrariness.  Note
  that the definitions in the present paper allow for a nonzero
  transverse component of ${\bf j}_{\mbox{\tiny{trans}}}$ in
  equilibrium for an inhomogeneous interacting electron system.}
\begin{equation}
  \label{eq:jtrans}
  \vec j_{\mbox{\scriptsize{trans}}} = \vec
  j_{\mbox{\scriptsize{total}}} - \vec j_{\mbox{\scriptsize{mag}}}
\end{equation}
which is the current of magnetized gauge transformed fermions, whereas
the magnetization current, as discussed above (see Eq.  \ref{eq:magn})
is the current associated with the attached current loops.

In addition, particles bound to magnetization should experience an
effective potential associated with any local changes in the magnetic
field.  Thus we define the effective scalar potential
\begin{equation}
  \label{eq:Aeff}
  A_0^{\mbox{\scriptsize{eff}}} = A_0 + \mum \delta B.
\end{equation}
This interaction of the bound magnetization with the magnetic field
should be thought of as the effective potential associated with the
local change in the cyclotron energy.

If we keep the conventions that all perturbations are applied with
$\vec q \| \hatn x$, and use the Coulomb gauge again, we can rewrite
Eqns. \ref{eq:jtrans} and \ref{eq:Aeff} as
\begin{eqnarray}
  j_{\mbox{\scriptsize{total}}} &=& M \label{eq:MJ}
  j_{\mbox{\scriptsize{trans}}} \\ A_{\mbox{\scriptsize{eff}}} &=&
  M^\dagger A \label{eq:MA}
\end{eqnarray}
where
\begin{equation}
  \label{eq:Mdef2}
  M = \left[ \begin{array}{cc} 1 & 0 \\ iq \mu_M & 1
\end{array} \right]_.
\end{equation}
In these equations, all currents are written as two vectors
$(j_0,j_y)$ and vector potentials are written as two vectors
($A_0,A_y$).  The matrix $M$ should be thought of as an operator that
attaches magnetization.  As discussed above, in the limit $m_b
\rightarrow 0$, we must have ${\mu_{\mbox{\tiny M}}} \rightarrow
\mu_b$ in the matrix $M$, but more generally we can allow corrections
on the interaction scale.  In the rest of this paper, however, we will
focus on the $m_b \rightarrow 0$ limit and consider ${\mu_{\mbox{\tiny
      M}}} = \mu_b$.

\subsection{Magnetized Modified RPA ($M^2 RPA$)}
\label{sec:M2RPA}

As discussed above, the (M)RPA approach does not properly model the
magnetization effects discussed in section \ref{sec:ZFR}.  This error
is presumably due to the fact that when we take the mass renormalized
mean field
solution as a starting point for a perturbation theory for the
Chern-Simons fermions, we lose the fact that the original electrons
travel in local cyclotron orbits.  In the approach discussed
here\cite{SimonStern}, we will recover this physics by artificially
attaching magnetization to each particle by hand.  This attachment is
not an exact transformation, but is rather a way of modeling behavior
that is lost when we take the mean field as a starting point.
However, as we will see below, within a Landau-Fermi liquid theory
picture, this attachment seems to give the correct quasiparticles for
the system.

The magnetized particles have the same interactions ($U$) as the
particles in the traditional Chern-Simons fermion picture.  However,
here, the magnetized fermions now respond to the effective potential
and the motion of these magnetized fermions yields only the transport
current response.  We thus define a matrix $\tilde K$ to be the {\it
  transport} current response of the electrons to the external {\it
  effective} potential.  In other words,
\begin{equation}
  K = M \tilde K M^\dagger.
\end{equation}
The `Magnetized Modified RPA' or $\mbox{M}^2$RPA is then defined by
setting $\tilde K$ equal to $K^{{\mbox{\tiny{MRPA}}}}$.  Thus we have
\begin{equation}
  \label{eq:M2def}
  K^{\M2} = M K^{{\mbox{\tiny{MRPA}}}} M^\dagger = M \left(
  [K^{0*}]^{-1} + {\cal F}_1 + U \right)^{-1} M^\dagger.
\end{equation}
It should be noted that
\begin{equation} K_{00}^{\M2} =
  K_{00}^{{\mbox{\tiny{MRPA}}}}
\end{equation}
and therefore the exact diagonalizations\cite{SH} that agreed well
with calculations of $K_{00}$ in the MRPA agree equally well with
predictions of the $\mbox{M}^2$RPA. However, the MRPA and
$\mbox{M}^2$RPA differ at finite $q$ in their predictions for the
other elements of the matrix $K$.  For example,
\begin{equation}
  \label{eq:K10M2}
  K_{10}^{\M2} = K_{10}^{{\mbox{\tiny{MRPA}}}} + i q \mum
  K_{00}^{{\mbox{\tiny{MRPA}}}} {}_.
\end{equation}
It should be noted however, that all finite $q$ experimental
tests\cite{WillettReview} of the Chern-Simons theory to date have
measured only $K_{00}$ and therefore do not distinguish between the
MRPA and the $\mbox{M}^2$RPA.  As required, in the limit $m_b
\rightarrow 0$, the $\mbox{M}^2$RPA correctly describes the static
response properties described above.  For example, Eq. \ref{eq:K10M2}
clearly satisfies Eq. \ref{eq:Klim}.

As is the case for the MRPA, we expect the $\mbox{M}^2$RPA, in
addition to describing the $\nu=\frac{1}{2m}$ Fermi liquid states,
should properly describe the Jain series of quantized states
$\nu=\frac{p}{2mp+1}$ for small $p$.  At large values of $p$, in the
case of Coulomb interactions, the description should be modified to
account for the effects of the singular infra-red gauge fluctuations.
In particular, the excitations at high $q$ are sensitive to the
infra-red divergence of the effective mass due to the gauge field
fluctuations\cite{Ady,Kim} which are neglected in $\mbox{M}^2$RPA.

\section{Fermi Liquid Theory}
\label{sec:FLT}

We now turn to discuss how the $\mbox{M}^2$RPA fits into the general
picture of a Fermi liquid theory of the $\nu=\frac{1}{2m}$ state.  In
essence, we will show that $\mbox{M}^2$RPA roughly amounts to adopting
the Fermi liquid picture of Ref.~\cite{Ady} as describing the dynamics
of magnetized composite fermion quasiparticles rather than
unmagnetized ones.

In Landau Fermi liquid theory for fermions with short ranged
interactions, such as ${}^3$He, the response function $K$ is given by
the solution of a Landau-Boltzmann equation\cite{Nozieres,Pines} which
describes the dynamics of quasiparticles near the Fermi surface.  In
such an approach, the quasiparticles are characterized by their
effective mass, $m^*$, and by the Landau interaction function,
$f({\vec k},{\vec k'})$, describing the {\it short range} interaction
between quasiparticles of momenta $\bf k$ and $\bf k'$.  In the case
of ${}^3$He, the quasiparticle effective mass is approximately three
times the bare mass, such that the quasiparticle is quite different
from the original particle.  In our composite fermion system, our
quasiparticle will not only have a renormalized mass, but also a
renormalized magnetization.

For fermions with long ranged interactions\cite{Nozieres,Pines}, the
Silin extension of the Landau theory asserts that it is the
polarization $\Pi$ that is described by the Landau-Boltzmann equation
(See Eq. \ref{eq:KPi}) rather than the full response $K$.  In other
words, Eq.~\ref{eq:KPi} separates out the Hartree part of the long
ranged interaction such that $\Pi$ gives the quasiparticle response to
the sum of the external vector potential and the induced internal
vector potential.  The Landau-Silin approach has been very successful
for the description of electrons in metals\cite{Nozieres,Pines,Lee}
(where there is only a long ranged Coulomb interaction and no
Chern-Simons interaction).  There, $\Pi$ is calculated with a
Boltzmann equation describing the dynamics of quasiparticles of mass
$m^*$ interacting via a {\it residual} short ranged interaction
$f({\vec k},{\vec k'})$.  Here we will try to construct a similar
Landau-Silin theory for the magnetized quasiparticles in the
Chern-Simons theory.

For the Chern-Simons theory, in addition to separating the long ranged
part of the interaction $U$, for the magnetized fermions, further
separation should be carried out to remove the magnetization effects.
To this end we define a response function $\tilde \Pi$ by
\begin{equation}
  \label{eq:tildePi}
  \Pi = M \tilde \Pi M^\dagger.
\end{equation}
By definition, $\tilde \Pi$ relates the transport current of the {\it
  magnetized} quasiparticles to the {\it effective} total vector
potential, including both external and internally induced
contributions (See Eqns. \ref{eq:MJ}, \ref{eq:MA} and \ref{eq:Atotal}).
For the Chern-Simons system it is $\tilde \Pi$ which we claim is given
by a Landau-Boltzmann equation describing the dynamics of
quasiparticles with the finite effective mass $m^*$ interacting via a
residual short ranged interaction $f({\vec k},{\vec k'})$.

\subsection{Boltzmann Transport}
\label{sec:Boltzmann}

In the Chern-Simons Fermi liquid, as in traditional Fermi liquid
theory, the (magnetized) quasiparticles are characterized by their
effective mass, $m^*$, and by the short ranged Landau interaction
function, $f({\vec k},{\vec k'})$. Since $|k| \approx |k'| \approx
\kf$, where $\kf$ is the Fermi momentum, $f$ is mostly\footnote{In the
  case of Coulomb or shorter ranged inter electron interactions,
  perturbative approaches\cite{Ady} find that $f$ may have a singular
  dependence on $|k|$.  One hopes that in a fully renormalized theory
  (nonperturbatively) these singularities do not prevent us from
  writing a Boltzmann transport equation.  We note that Kim et
  al\cite{Kim} recently showed that a form of Quantum Boltzmann
  equation can be derived independent of these singularities.} a
function of $\theta$, the angle between $\vec k$ and $\vec k'$.  It is
often more convenient to work with the Fourier transformed quantity
\begin{equation}
  \label{eq:fldef}
  f_l = \frac{1}{2 \pi} \int_0^{2 \pi} d \theta f(\theta) e^{i l
    \theta}
\end{equation}
Due to the symmetry of the interaction function $f(\theta) =
f(2\pi-\theta)$ we expect that $f_l = f_{-l}$.

In order to calculate the response function $\tilde \Pi$, we keep with
the convention that the driving force $\vec F$ is applied with
wavevector $\vec q \| \hatn x$, and at frequency $\omega$ (i.e., the
perturbation is proportional to $e^{iqx-i\omega t }$). Writing the
fluctuations of the Fermi surface as $\delta n(\vec p) = \nu(\theta)
\delta(|\vec p| - \pf)$ where $\theta$ is the direction of $\vec p$ on
the Fermi surface\footnote{The definition of $\nu$ agrees with that in
  references \cite{Ady}, \cite{Nozieres}, and \cite{Pines} but differs
  from the function $f$ used in references \cite{Lee} and
  \cite{Simonhalp} by a factor of $\vf$.}, the Boltzmann transport
equation can be written as\cite{Nozieres,Pines,Lee}
\begin{equation}
\label{eq:kin1}
-i\omega \nu(\theta) + iq\vf \cos(\theta) [\nu(\theta) + \delta
\epsilon_1(\theta)] = \vec F \cdot \hatn{n}(\theta)
\end{equation}
where $\vf = \pf/m^*$ is the mass renormalized Fermi velocity.
\begin{equation}
  \label{eq:eps1}
  \delta \epsilon_1(\theta) = \frac{m^*}{(2 \pi \hbar)^2} \int
  d\theta' f(\theta - \theta') \nu(\theta'),
\end{equation}
and the directional  vector is given by
\begin{equation}
  \label{eq:hatntheta}
  \hatn n(\theta) = (\cos \theta, \sin \theta).
\end{equation}
Equation \ref{eq:kin1} is just the usual Boltzmann equation of Fermi
liquid theory.  However, here the driving force is given by the total
{\it effective} electric field
\begin{equation}
  \vec F = -e \vec
  E_{\mbox{\scriptsize{eff}}}^{\mbox{\scriptsize{total}}} = -e \left(
  \nabla A_{\mbox{\scriptsize{eff}}\;0}^{\mbox{\scriptsize{total}}} -
  \frac{d}{dt} \vec
  A_{\mbox{\scriptsize{eff}}}^{\mbox{\scriptsize{total}}} \right)_.
\end{equation}
where (See Eqs. \ref{eq:Atotal} and Eq. \ref{eq:MA})
\begin{equation}
   A_{\mbox{\scriptsize{eff}}}^{\mbox{\scriptsize{total}}} = M^\dagger
   A^{\mbox{\scriptsize{total}}}.
\end{equation}

Once one has solved Eq. \ref{eq:kin1} for $\nu(\theta)$, The local
charge density can be written as the density of quasiparticles
\cite{Nozieres,Pines,Lee}
\begin{equation}
  \label{eq:j0theta}
  j_0 = \frac{-e \pf}{(2 \pi \hbar)^2} \int d \theta \nu(\theta).
\end{equation}
Similarly, the motion of these magnetized quasiparticles gives the local {\it
  transport} current density
\begin{eqnarray}
  \label{eq:jtheta}
  \vec j_{\mbox{\scriptsize{trans}}} &=& \label{eq:current1}
  \frac{1}{m^*} \left[ \frac{-e \pf^2}{(2 \pi \hbar)^2}\right] \int d
  \theta {\hatn n}(\theta) \left\{ \nu(\theta) + \delta
  \epsilon_1(\theta) \right\} \\ &=& \frac{1}{m_b} \left[ \frac{-e
    n_e}{\pi}\right] \int d \theta \hatn n(\theta) \nu(\theta).
\end{eqnarray}
Thus one can easily find the magnetized quasiparticle resistivity
matrix $\tilde \rho_{cf}$ relating the effective electric field to the
transport current via the 2 $\times$ 2 matrix equation (cf. Eq.
\ref{eq:ejcf})
\begin{equation}
  \vec E_{\mbox{\scriptsize{eff}}}^{\mbox{\scriptsize{total}}} =
  \tilde \rho_{cf} \vec j_{\mbox{\scriptsize{trans}}}.
\end{equation}
The response matrix $\tilde \Pi$ is then given by (cf. Eq.
\ref{eq:Pirho})
\begin{equation}
  \label{eq:Pirhocf}
  \tilde \Pi = [T \tilde \rho_{cf} T]^{-1}
\end{equation}

We now have a prescription for calculating the response $K$ of the
Chern-Simons Fermi liquid given the effective mass $m^*$ and the
interaction function $f(\theta)$.  To reiterate, the prescription is
to solve the Boltzmann equation (Eq. \ref{eq:kin1}) for $\nu(\theta)$
and calculate the current using Eq. \ref{eq:jtheta} to get the
magnetized composite fermion resistivity $\tilde \rho_{cf}$.  The
response $K$ can then be obtained by using Eqns. \ref{eq:Pirhocf},
\ref{eq:tildePi} and \ref{eq:KPi}.

\subsection{Separating Singular Fermi Liquid Coefficients}
\label{sec:sep}

As discussed above, one expects that the effective mass, which
determines the energy scale of the low energy excitations, should be
set by the Coulomb interaction scale.  Similarly, one
expects\cite{Ady} that the interaction function $f(\theta)$ should be
on the interaction scale (ie, proportional to $1/m^*$).  However, two
important restrictions on $f$ yield pieces of $f$ that are set by the
larger scale $1/m_b$.

A well known result of Fermi liquid theory\cite{Nozieres,Pines} is
that the Fermi liquid coefficients $f_0$ and $f_1$ are fixed by the
identities
\begin{eqnarray}
  \frac{1}{m_b} &=& \frac{1}{m^*} + \frac{f_1}{2 \pi \hbar^2}.
  \label{eq:f1effmass} \\
  \frac{d \mu}{d n} &=& \frac{2 \pi \hbar^2}{m^*} + f_0
  \label{eq:f0comp}
\end{eqnarray}
The identity \ref{eq:f1effmass} is a result of Galilean
invariance\cite{Nozieres,Pines}.  (Note that $f_1$ refers to the first
Fourier mode of excitations of the Fermi surface which corresponds to
a Galilean boost).  Thus, $f_1$ is clearly on the larger scale $1/m_b$
rather than the interaction scale.  Furthermore, we claim that the sum
rule \ref{eq:f0comp} fixes $f_0$ to be on the scale $1/m_b$ also.
This counterintuitive result is due to the fact that the
compressibility derivative $\frac{d \mu}{d n}$ is taken at fixed
$\Delta B$.  One can understand this\cite{Ady,Kim} by realizing that
the Fermi liquid theory uses the mean field zero effective field
solution for its ground state.  When a particle is added or
subtracted, in order to maintain a Fermi liquid (ie, zero effective
field), the external field must increased by $\tilde \phi$ flux quanta
to compensate for the added Chern-Simons field.  Thus, at fixed
$\Delta B=0$, the magnetic field is linked to the density $n$ via $B =
\tilde \phi n \Phi_0$.  In the limit $m_b \rightarrow 0$, the
interaction energy between the magnetization ${\bf M} = \mu_b n$ and
the external field is given by
\begin{equation}    E = {\bf M} \cdot {\bf B} = \frac{\pi \tilde \phi
    \hbar^2 n^2}{m_b}.
\end{equation}
Of course this can also be thought of as the cyclotron energy.
Differentiating this with respect to $n$ we obtain a magnetization
contribution to the chemical potential
\begin{equation}
  \mu^{\mbox{\scriptsize{mag}}} = \frac{2 \pi \tilde \phi \hbar^2 n}{m_b}=
  \hbar \omega_c
\end{equation}
such that the magnetization contribution $\tilde f_0$ to the zeroth
Fermi liquid coefficient $f_0$ is given by
\begin{equation}
  \tilde f_0 = \frac{d \mu^{\mbox{\scriptsize{mag}}}}{d n} = \frac{2
    \pi \tilde \phi \hbar^2}{m_b}
\end{equation}
which is also the inverse compressibility of free electrons of mass
$m_b$ at constant $\Delta B$.  The coefficient $f_0$ is written $f_0 =
\tilde f_0 + \delta f_0$ where $\tilde f_0$ is ${\cal O}(m_b^{-1})$
and $\delta f_0$ is on the smaller interaction scale.  As mentioned in
Ref.~\cite{Ady}, in the limit $m_b \rightarrow 0$, the requirement
that the low energy spectrum is independent of $m_b$ forces the other
interaction coefficients ($f_l$ for $l \ne 0,1$) to be on the
interaction scale.  In addition we note that using the Pauli exclusion
principle a sum rule can be derived for the remaining Fermi liquid
coefficients $f_l$ for $l \ne 0,1$.  This sum rule is derived
explicitly in Appendix A.

Since in the limit of $m_b \rightarrow 0$, $\tilde f_0$ and $f_1$ are
on the bare mass scale whereas all other coefficients $f_l$ (as well
as $\delta f_0$) are expected to be on the smaller interaction scale,
we will separate out the contributions of these two coefficients by
writing
\begin{equation}
  \label{eq:corrs}
  \tilde \Pi^{-1} = [\tilde \Pi^*]^{-1} + \tilde {\cal F}_0 + {\cal
    F}_1
\end{equation}
where
\begin{equation}
  \label{eq:F0def} \tilde {\cal F}_0 =
  \left(
  \begin{array}{cc}  \tilde f_0 & 0 \\ 0 & 0 \end{array}
  \right)_.
\end{equation}
and ${\cal F}_1$ is given by Eq. \ref{eq:F1def}.  The function $\tilde
\Pi^*$ is to be calculated using a Landau-Boltzmann equation
representing quasiparticles with the same effective mass $m^*$ and
interaction coefficients $f_l$ except that $f_1$ is artificially set
to zero and the magnetic contribution $\tilde f_0$ is subtracted off
of $f_0$.  Once again, the form of Eq. \ref{eq:corrs} looks like the
form of Eq. \ref{eq:KPi} where we have separated two interaction terms
and defined the remaining response $\tilde \Pi^*$ to be the response
of a similar Fermi liquid with those interactions removed.  The
separation of the coefficient $f_0$, analogous to taking $v(q)
\rightarrow v(q) + f_0$ in Eq.~\ref{eq:KPi}, is justified by noting
that $f_0$ corresponds to a short ranged density-density interaction.
Similarly, the separation of the coefficient $f_1$ is achieved by
noting that the the $f_1$ coefficient corresponds to a current-current
interaction (${\cal F}_1$) which can similarly be added on in Eq.
\ref{eq:corrs}.  The separation of the nonzero $f_1$
coefficient\cite{Simonhalp} is analogous to that described in
Eq.~\ref{eq:Pi*} (the coefficient of the matrix in Eq.  \ref{eq:F1def}
is proportional to $f_1$) and is derived explicitly in
Ref.~\cite{Simonhalp}.  Note that the separation of the effects of
Fermi liquid coefficients by treating them as density-density and
current-current interactions can only be done for $f_0$ and $f_1$ and
not for any $f_l$ for $l > 1$.  Having made this separation, we expect
that the response $\tilde \Pi^*(q,\omega)$ is independent of $m_b$ in
the limit $m_b \rightarrow 0$ and is well behaved for all values of
$q/m_b$.  The transformation Eqns.~\ref{eq:KPi}, \ref{eq:Pi*},
\ref{eq:F1def}, \ref{eq:tildePi}, and \ref{eq:corrs} do not in
themselves involve any approximations, and may be considered simply as
a means of defining a new `irreducible' response function $\tilde
\Pi^*(q, \omega)$.

\subsection{Relation to $\mbox{M}^2$RPA}
\label{sec:relation}

To relate this Fermi liquid approach to the $\mbox{M}^2$RPA we note
the identity
\begin{equation}
  \label{eq:Uiden}
  U + \tilde {\cal F}_0 = M^\dagger{}^{-1} U M^{-1}
\end{equation}
which holds in the limit $m_b \rightarrow 0$.  This identity is a
statement of the fact that if you allow the magnetization to see the
Chern-Simons magnetic field as well as the external magnetic field,
then the $1/m_b$ contribution to $f_0$ will vanish since the
magnetization now sees zero magnetic field on average.  We will also
need
\begin{equation}
  \tilde {\cal F}_0 = M^{\dagger} \tilde {\cal F}_0 M,
\end{equation}
which is just the statement that a density-density interaction does
not care whether or not the particles are magnetized.  Using these
identities, we find that $\mbox{M}^2$RPA defined in Eq. \ref{eq:M2def}
is equivalent to approximating $\Pi^*$ by $K^{0*}$, the response of a
free Fermi gas of particles of mass $m^*$, and calculating the
response using Eqns.~\ref{eq:KPi}, \ref{eq:tildePi}, and
\ref{eq:corrs}.

We note that in Fermi liquid theory, the Landau-Boltzmann equation
does not correctly describe the Landau diamagnetic contribution to the
transverse static response.  Similarly, we suspect that here the
function $\tilde \Pi^*_{11}$ derived from the Landau-Boltzmann
equation lacks a term of the form $q^2 \chi$ where $\chi$ is some
appropriate Landau susceptibility which we expect to be on the scale
of the interaction strength. As usual, if we fix the ratio $\omega/q$
to be nonzero, and take $q \rightarrow 0$, this diamagnetic term
becomes negligible.  However, when $\tilde \Pi^*$ is approximated as
$K^{0*}$ for the $\mbox{M}^2$RPA, this diamagnetic contribution is
included at least approximately.

\section{Further Comments and Conclusions}
\label{sec:conclusion}

\subsection{Effect of Other Fermi Liquid Coefficients}

Clearly, the $\mbox{M}^2$RPA involves neglecting Fermi liquid
coefficients $f_l$ for $l \ne 0,1$.  Although this formally violates
the sum rule of appendix A, the neglect of these interaction terms is
probably quite reasonable.  In previously studied Fermi liquid
theories (Helium--3 and electrons in metals) although the first few
Fermi liquid coefficients may be large, the higher ones become rapidly
smaller\cite{Pines}.

To elucidate the effects of additional nonzero Fermi liquid
coefficients, we consider the addition of a nonzero magnetic field
$\Delta B$.  At the Jain series of filling fractions
$\frac{p}{2mp+1}$, the composite fermions fill precisely $p$ Landau
levels, resulting in fractionally quantized states.  The Boltzmann
excitation spectrum for composite fermions for these
states\cite{Simonhalp,Lee} is given by
\begin{equation}
  \label{eq:highermodes}
  \omega_n = n \left( 1 + \frac{m^* f_n}{2 \pi \hbar^2}\right) \Delta
  \omega_c^*
\end{equation}
where $\Delta \omega_c^* = e \Delta B/m^*$, and $n$ is a positive
integer.  The residue (or weight) of the $n^{th}$ excitation mode is
proportional to $q^{2n}$.  Since only the $n=1$ mode has weight in the
small $q$ limit, this is the only mode that is altered by the
Chern-Simons or Coulomb interactions (Eq. \ref{eq:KPi}).  Thus, the
response spectrum (ie, location of poles of $K_{00}$) is identical to
the composite fermion spectrum predicted by Eq. \ref{eq:highermodes}
except that the $n=1$ mode is pushed up to the cyclotron frequency
(see Eq.  \ref{eq:f1effmass}) as required by Kohn's
theorem\cite{Lopez,Simonhalp}.

The $q \rightarrow 0$ spectrum shown by Eq. \ref{eq:highermodes} would
suggest that it would be very easy to extract the value of the Fermi
liquid coefficients $f_l$ from the response of the system.  However,
we point out that the spectrum predicted by the above Fermi liquid
theory (or by the MRPA and $\mbox{M}^2$RPA) yields a spectrum of
single quasiparticle excitations only.  This single particle
excitation spectrum should be correct at low frequency, but at higher
frequency one can create multiple low energy excitations.  At least
under some conditions, at finite $\Delta B$, these multiple
excitations may have more weight than the single particle excitations
in the $q \rightarrow 0$ limit\cite{Song}, making it more difficult to
accurately extract Fermi liquid coefficients directly from an excitation
spectrum using Eq. \ref{eq:highermodes}.

\subsection{Connection with Other Recent Work}

Using the $\mbox{M}^2$RPA approach, we can calculate the response $K$
and hence $\Pi$ (Eq. \ref{eq:KPi}) and hence the composite fermion
conductivity $\sigma_{cf} = [\rho_{cf}]^{-1}$ via Eq. \ref{eq:Pirho}.
This could equivalently be calculated by using Eqns. \ref{eq:corrs}
and \ref{eq:tildePi} along with the approximation $\tilde \Pi^* =
K^{0*}$ which, as discussed above, is equivalent to the
$\mbox{M}^2$RPA.  Either approach yields the limiting low frequency
and wavevector composite fermion Hall conductivity for small $m_b$
\begin{equation}
  \lim_{q \rightarrow 0} \lim_{\omega \rightarrow 0}
  [\sigma_{cf}]_{xy} = -\left(\frac{1}{2 \tilde \phi} \right)
\left(\frac{e^2}{2
    \pi \hbar} \right).
\end{equation}
For $\nu =\frac{1}{2}$, this is precisely half the value found in Ref.
\cite{DHLee} {\it in the opposite order of limit and in the presence of
  disorder}.  Since these two results are for slighlty different cases, it
is not clear that there is any contradiction.  Ref.  \cite{DHLee} also
calculates the above order of limits for a clean system and finds it to be
zero to first order in perturbation theory in $\tilde \phi$.  Although the
calculations described in Ref.  \cite{DHLee} would be a natural direction
for attempting to understand this attachment, at lowest order the
magnetization effects are not seen.  Once again this result does not
directly contradict our work since it is only perturbative.  Note that our
result, being inversely proportional to $\tilde \phi$, may indicate why
the perturbative approach yields zero.

\subsection{Possible Relation to Other Pictures of Quantum Hall States}

Since much of our knowledge of Quantum Hall states stems from the use
of trial wavefunctions\cite{Jain,Prange,Haldane}, it is natural to try
to make contact with these approaches.  Typically the trial
wavefunctions are projected to the lowest Landau level which in some
senses can be thought of as the $m_b \rightarrow 0$
limit\footnote{Note that projection and $m_b \rightarrow 0$ are not
  formally equivalent.  See for example Reference \cite{Sondhi}.}.
Since the magnetization attachment described in this paper is
concerned with exactly this limit, it is interesting to see to what
extent the physics described in this paper matches the physics
described by the trial wavefunctions.  Particularly interesting would
be a comparison to the predictions of the $\nu=\frac{1}{2}$ lowest
Landau level wavefunction constructed by Haldane\cite{Haldane}.

\subsection{Conclusions}

The $\mbox{M}^2$RPA, describes the $\nu=\frac{1}{2m}$ state as a Fermi
liquid of magnetized composite fermions with a finite renormalized
effective mass $m^*$, an $f_1$ parameter dictated by Galilean
invariance and an $f_0$ parameter originating from the interaction of
the magnetization with the magnetic field.  All remaining Fermi liquid
parameters (which are expected to be on the much smaller interaction
scale) are neglected.  The $\mbox{M}^2$RPA predicts the same $K_{00}$
as the MRPA, but in contrast it yields the correct behavior for
$K_{01}, K_{10}$, and $K_{11}$ in the limit $m_b \rightarrow 0$ for
arbitrary small $q$.

By separating the pieces due to magnetization and due to singular
Fermi liquid coefficients (in the $m_b \rightarrow 0$ limit) we
identify a response function $\tilde \Pi^*$ that is represented by the
solution of a well behaved Landau-Boltzmann equation (up to
diamagnetic terms).  Our claim that $\tilde \Pi^*$ is well behaved in
the limit $m_b \rightarrow 0$ we believe to be an exact statement
(although we have not proved it rigorously) independent of the
approximation used to define the $\mbox{M}^2$RPA.

\ack{Much of the work described in this paper was developed in
  collaboration with Ady Stern and Bertrand Halperin.  It is a
  pleasure to acknowledge this fruitful collaboration.  I would also
  like to thank P.~A.~Lee, Y.-B. Kim, S. Kivelson, and A.~Castro-Neto
  for helpful discussions.  The author would also like to thank the
  ITP in Santa Barbara for its hospitality.  This work was supported
  by NSF Grants No. DMR-94-16910 and DMR-95-23361}.

\appendix
\section{Fermi Liquid Sum Rule}
\label{app:sumrule}

By using the Pauli exclusion principle for the forward scattering
amplitude, a sum rule can be derived for the Landau coefficients for a
Fermi liquid\cite{LandauII}.  In two dimensions for a spinless Fermi
liquid, the form of this sum rule is (In this appendix $\hbar = e = c
= 1$).
\begin{equation}
  \label{eq:appzero}
  \sum_{l=-\infty}^{\infty} \frac{f_l}{(2 \pi/m^*) + f_l} = 0
\end{equation}
In Reference \cite{Platzman} this sum rule is generalized to the
case of Landau-Silin Fermi liquid theory for systems with long ranged
Coulomb interactions.  In two dimensions, for a spinless Fermi liquid
with long range interactions, the sum rule then becomes
\begin{equation}
  \label{eq:appone}
  \sum_{\stackrel{l=-\infty}{l \ne 0}}^{\infty} \frac{f_l}{(2 \pi/m^*)
    + f_l} = -1
\end{equation}

In this appendix we will derive the form of the sum rule in two
dimensions for a spinless two dimensional Fermi liquid interacting via
a long ranged Chern-Simons gauge field as well as via a direct
`Coulomb' interaction $v(r)$.  As much as possible, we will use the
notations of references \cite{Nozieres} and \cite{Platzman}.  In this
derivation, we will assume for simplicity that there are no
complications due to infra-red divergent gauge field fluctuations.
This should be rigorously true in the case where the interaction is
longer ranged than Coulomb.  One hopes that in the case of Coulomb and
shorter ranged interactions, cancelation of divergences similar to
those found in the calculations of the electromagnetic
response\cite{Kim,Millis,Kvesh} will lead to a fully renormalized
theory that also obeys the sum rule derived here.  Note that a formal
derivation of the sum rule is given first, followed by a simple
phenomenological interpretation.

As in reference \cite{Platzman} we write the full vertex
function ${}^0\Gamma$ in terms of 'proper' or irreducible four point
function $\tilde \Gamma$ by writing
\begin{equation} \nonumber
  {}^0\Gamma(p,p';\bar \omega) =   {}^0\tilde \Gamma(p,p';\bar \omega)
    + \tilde \Lambda(p,\bar \omega)
     \left[1 - U(q) \tilde
   S(\bar \omega) \right]^{-1}  U(q) \tilde
   \Lambda(p',\bar \omega), \label{eq:bubblesum}
\end{equation}
where $p = (\omega, \vec k)$, $p' = (\omega',\vec k')$ and $\bar
\omega = (\epsilon,\vec q) = p' - p$.  Note that all three vectors
will be arranged such that the zeroth element of the vector $p$ is the
frequency element (This differs from the notation of references
\cite{Nozieres} and \cite{Platzman}).  Here, the
three-vector $\tilde \Lambda$ is the proper three point vertex
function, and the three by three matrix $\tilde S$ is the proper
polarization propegator.  Also, we have the interaction matrix
\begin{equation}
  \label{eq:Uofq}
  U(q) = \frac{-i}{2 \pi} \left( \begin{array}{ccc} v(q) & 0 & ic(q) \\
  0 & 0 & 0 \\ -i c(q) & 0 & 0 \end{array} \right)
\end{equation}
with $ c(q) = 2 \pi \tilde \phi/q$ the Chern-Simons gauge
interaction\cite{HLR}, and $v(q)$ the direct Coulomb interaction.  As
mentioned above, although the physical case is $v(q) = 2 \pi/
(\epsilon q)$, we may want to consider other functional forms.  Note
that the zeroth row and column of the matrix $U$ represent the
interactions of the density whereas the first and second row and
column represent the longitudinal and transverse current respectively.
Equation \ref{eq:bubblesum} is shown diagramatically in Figure
\ref{fig:diagram}.

\begin{figure}[htbp]
  \begin{center}
    \leavevmode
\setlength{\unitlength}{0.00041700in}%
\begingroup\makeatletter\ifx\SetFigFont\undefined
\def\x#1#2#3#4#5#6#7\relax{\def\x{#1#2#3#4#5#6}}%
\expandafter\x\fmtname xxxxxx\relax \def\y{splain}%
\ifx\x\y   
\gdef\SetFigFont#1#2#3{%
  \ifnum #1<17\tiny\else \ifnum #1<20\small\else
  \ifnum #1<24\normalsize\else \ifnum #1<29\large\else
  \ifnum #1<34\Large\else \ifnum #1<41\LARGE\else
     \huge\fi\fi\fi\fi\fi\fi
  \csname #3\endcsname}%
\else
\gdef\SetFigFont#1#2#3{\begingroup
  \count@#1\relax \ifnum 25<\count@\count@25\fi
  \def\x{\endgroup\@setsize\SetFigFont{#2pt}}%
  \expandafter\x
    \csname \romannumeral\the\count@ pt\expandafter\endcsname
    \csname @\romannumeral\the\count@ pt\endcsname
  \csname #3\endcsname}%
\fi
\fi\endgroup
\begin{picture}(8572,2744)(279,-2483)
  \thicklines \put(8101,-1111){\oval(200,400)} \put(4426,-1561){\line(
    1,-1){600}} \put(4426,-961){\line( 1, 1){600}}
  \put(3826,-1561){\line(-1,-1){600}} \put(3826,-961){\line(-1,
    1){600}}
  \multiput(3451,-586)(12.50000,-12.50000){7}
  {\makebox(13.3333,20.0000){\SetFigFont{7}{8.4}{rm}.}}
  \put(3451,-586){\vector(-1, 1){0}}
  \multiput(4801,-1936)(-12.50000,12.50000){7}
  {\makebox(13.3333,20.0000){\SetFigFont{7}{8.4}{rm}.}}
  \put(4801,-1936){\vector( 1,-1){0}}
  \multiput(3601,-1786)(-12.50000,-12.50000){7}
  {\makebox(13.3333,20.0000){\SetFigFont{7}{8.4}{rm}.}}
  \put(3601,-1786){\vector( 1, 1){0}}
  \multiput(4651,-736)(12.50000,12.50000){7}
  {\makebox(13.3333,20.0000){\SetFigFont{7}{8.4}{rm}.}}
  \put(4651,-736){\vector(-1,-1){0}} \put(5701,239){\line( 1,-1){600}}
  \put(6901,239){\line(-1,-1){600}} \put(6001,-61){\line( 1, 0){600}}
  \put(6301,-1861){\line(-1,-1){600}} \put(6301,-1861){\line(
    1,-1){600}} \put(6001,-2161){\line( 1, 0){600}}
  \multiput(6676,14)(12.50000,12.50000){7}
  {\makebox(13.3333,20.0000){\SetFigFont{7}{8.4}{rm}.}}
  \put(6676, 14){\vector(-1,-1){0}}
  \multiput(5926,-2236)(-12.50000,-12.50000){7}
  {\makebox(13.3333,20.0000){\SetFigFont{7}{8.4}{rm}.}}
  \put(5926,-2236){\vector( 1, 1){0}} \put(7501,239){\line(
    1,-1){600}} \put(8701,239){\line(-1,-1){600}}
  \put(7801,-61){\line( 1, 0){600}}
  \put(8101,-1861){\line(-1,-1){600}} \put(8101,-1861){\line(
    1,-1){600}} \put(7801,-2161){\line( 1, 0){600}} \multiput(8476,
  14)(12.50000,12.50000){7}
  {\makebox(13.3333,20.0000){\SetFigFont{7}{8.4}{rm}.}}
  \put(8476, 14){\vector(-1,-1){0}}
  \multiput(7726,-2236)(-12.50000,-12.50000){7}
  {\makebox(13.3333,20.0000){\SetFigFont{7}{8.4}{rm}.}}
  \put(7726,-2236){\vector( 1, 1){0}}
  \multiput(5776,164)(12.50000,-12.50000){7}
  {\makebox(13.3333,20.0000){\SetFigFont{7}{8.4}{rm}.}}
  \put(5776,164){\vector(-1, 1){0}}
  \multiput(6826,-2386)(-12.50000,12.50000){7}
  {\makebox(13.3333,20.0000){\SetFigFont{7}{8.4}{rm}.}}
  \put(6826,-2386){\vector( 1,-1){0}}
  \multiput(7576,164)(12.50000,-12.50000){7}
  {\makebox(13.3333,20.0000){\SetFigFont{7}{8.4}{rm}.}}
  \put(7576,164){\vector(-1, 1){0}}
  \multiput(8626,-2386)(-12.50000,12.50000){7}
  {\makebox(13.3333,20.0000){\SetFigFont{7}{8.4}{rm}.}}
  \put(8626,-2386){\vector( 1,-1){0}}
  \put(3826,-1561){\framebox(600,600){}} \put(6301,-586){\line( 0,
    1){150}} \put(6301,-886){\line( 0, 1){150}}
  \put(6301,-1186){\line( 0, 1){150}} \put(6301,-1486){\line( 0,
    1){150}} \put(6301,-1786){\line( 0, 1){150}}
  \put(8101,-1786){\line( 0, 1){150}} \put(8101,-586){\line( 0,
    1){150}} \put(8101,-886){\line( 0, 1){150}}
  \put(8101,-1486){\line( 0, 1){150}} \put(1501,-1561){\line(
    1,-1){600}} \put(1501,-961){\line( 1, 1){600}}
  \put(901,-1561){\line(-1,-1){600}} \put(901,-961){\line(-1, 1){600}}
  \multiput(526,-586)(12.50000,-12.50000){7}
  {\makebox(13.3333,20.0000){\SetFigFont{7}{8.4}{rm}.}}
  \put(526,-586){\vector(-1, 1){0}}
  \multiput(1876,-1936)(-12.50000,12.50000){7}
  {\makebox(13.3333,20.0000){\SetFigFont{7}{8.4}{rm}.}}
  \put(1876,-1936){\vector( 1,-1){0}}
  \multiput(1726,-736)(12.50000,12.50000){7}
  {\makebox(13.3333,20.0000){\SetFigFont{7}{8.4}{rm}.}}
  \put(1726,-736){\vector(-1,-1){0}}
  \multiput(676,-1786)(-12.50000,-12.50000){7}
  {\makebox(13.3333,20.0000){\SetFigFont{7}{8.4}{rm}.}}
  \put(676,-1786){\vector( 1, 1){0}}
  \put(901,-1561){\framebox(600,600){}} \put(976,-961){\line(
    0,-1){600}} \put(1051,-961){\line( 0,-1){600}}
  \put(1126,-961){\line( 0,-1){600}} \put(1201,-961){\line(
    0,-1){600}} \put(1276,-961){\line( 0,-1){600}}
  \put(1351,-961){\line( 0,-1){600}} \put(1426,-961){\line(
    0,-1){600}} \put(901,-1036){\line( 1, 0){600}}
  \put(901,-1111){\line( 1, 0){600}} \put(901,-1186){\line( 1,
    0){600}} \put(901,-1261){\line( 1, 0){600}} \put(901,-1336){\line(
    1, 0){600}} \put(901,-1411){\line( 1, 0){600}}
  \put(901,-1486){\line( 1, 0){600}}
  \put(2476,-1336){\makebox(0,0)[lb]{\smash{\SetFigFont{10}{12.0}{rm}=}}}
  \put(5251,-1336){\makebox(0,0)[lb]{\smash{\SetFigFont{10}{12.0}{rm}+
        }}}
  \put(7201,-1336){\makebox(0,0)[lb]{\smash{\SetFigFont{10}{12.0}{rm}+
        }}}
  \put(8851,-1336){\makebox(0,0)[lb]{\smash{\SetFigFont{10}{12.0}{rm}+
        ...}}}
\end{picture}
\hspace{.5in} $ {}^0\Gamma(p,p';\bar \omega) = {}^0\tilde
\Gamma(p,p';\bar \omega) + \mbox{Self Consistent Field Terms} $
    \end{center}
  \caption{Separation of the four point function ${}^0\Gamma$ into its
    irreducible part ${}^0 \tilde \Gamma$ and self consistent field
    interaction contributions.  This is the diagramatic representation
    of Eq.  \protect{\ref{eq:bubblesum}}.  Here the dotted line is the
    interaction propegator $U$ that includes both Coulomb and
    Chern-Simons terms.}
  \label{fig:diagram}
\end{figure}

The Landau interaction function for a Landau-Silin Fermi liquid theory
is given in terms of the proper four point function
by\cite{Nozieres,Platzman}
\begin{eqnarray}
  f(\vec k, \vec k') &=& 2 \pi i z_k z_{k'}
\lim_{q/\epsilon
    \rightarrow 0} \lim_{\bar \omega \rightarrow 0} {}^0 \tilde \Gamma(p,p';
  \bar \omega)  \\
  &=& 2 \pi i z_k z_{k'} {}^0 \tilde \Gamma^0(\vec k, \vec k')
\end{eqnarray}
where $k$ and $k'$ are taken on the Fermi surface, and $z_k$ is the
quasiparticle renormalization.

The Pauli principle, on the other hand, dictates that\cite{Platzman}
\begin{equation}
  \lim_{\bar \omega \rightarrow 0} \lim_{(q/\epsilon) \rightarrow \infty}
  {}^0 \Gamma(p, p'; p - p') = {}^0 \Gamma^\infty(\vec k,\vec k') = 0
\end{equation}
Applying the same limits to equation \ref{eq:bubblesum} we obtain
\begin{eqnarray}
  \label{eq:bubble2}
  &{}&^0\tilde \Gamma^\infty(\vec k,\vec k) = \lim_{\bar \omega
    \rightarrow 0} \lim_{(q/\epsilon)\rightarrow \infty} {}^0\tilde
  \Gamma(p,p';\bar \omega) =
   \\ & & \nonumber   \tilde \Lambda^\infty(p,\bar \omega)
   \left[1 + U(q) \tilde S^\infty(\bar \omega) \right]^{-1} U(q)
     \tilde \Lambda^\infty(p,\bar \omega).
     \end{eqnarray}
where $\tilde \Lambda^\infty$ and $\tilde S^\infty$ are the
corresponding limits of $\tilde \Lambda$ and $\tilde S$.

Ward identities\cite{Nozieres} can be invoked to yield
\begin{equation}
  \tilde \Lambda^\infty(p, \bar \omega) = \frac{\vf}{z_k} \left(
  \begin{array}{c}
  \partial \kf/\partial \mu \\ 0 \\ 1 \end{array} \right)
\end{equation}
where $\vf = \kf/m^*$ is the Fermi velocity, and $\mu$ is the chemical
potential.  It should be noted that the longitudinal current element
vanishes because in the limit that $\bar{\omega} \rightarrow 0$ and $
(q/\epsilon) \rightarrow \infty$ we must have $q \perp k$.  Ward
identities can also be used to calculate the matrix \cite{Nozieres}
\begin{equation}
  \tilde S^\infty = - 2 \pi i \,\,\mbox{diag} \left[ \frac{\partial
    n}{\partial \mu}, \frac{n}{m_b}, \frac{n}{m_b} \right]
\end{equation}
where $m_b$ is the bare band mass and $n$ is the density.  Using these
relations in equation \ref{eq:bubble2} we find
\begin{eqnarray}
  g(\vec k, \vec k) &=&  2 \pi i z_k z_{k} {}^0\tilde \Gamma^\infty(\vec
  k,\vec k) \\ &=&  - {\vf}^2 \left[ (\partial \kf/\partial \mu)^2 /
  (\partial n/\partial \mu) + m_b/n \right].
\end{eqnarray}
Note that this relation holds for all interactions $v(q) \sim
q^{-\alpha}$ with $\alpha < 2$.

Using the relation\cite{Nozieres,Platzman}
\begin{equation}
  g(\vec k, \vec {k'}) = f(\vec k, \vec {k'}) - \int \frac{d\vec
    k''}{(2 \pi)^2} f(\vec k,\vec k'') g(\vec k'', \vec k') \delta(
  \epsilon_{k''} - \mu)
\end{equation}
we derive
\begin{equation}
  g_l = \frac{1}{2 \pi} \int_0^{2 \pi} d\theta g(\theta) e^{i l
    \theta} = \frac{ f_l}{ (2 \pi/m^*) + f_l}
\end{equation}
and thus
\begin{eqnarray}
  g(\theta = 0) &=& -{\vf}^2 \left[ (\partial \kf/\partial \mu)^2
  (\partial n/\partial \mu) + m_b/n \right] \\ & = & \sum_{l =
    -\infty}^{\infty} \frac{f_l}{(2 \pi/m^*) + f_l}
\end{eqnarray}
Finally, using the identities \ref{eq:f0comp} and \ref{eq:f1effmass}
along with ${\kf}^2 = 4 \pi n$ and $\vf = \kf/m^*$ we obtain
\begin{equation}
  \label{eq:appresult}
  \sum_{\stackrel{l = -\infty}{l \ne 0,-1,1}}^{\infty} \frac{f_l}{(2
    \pi/m^*) + f_l} = - 3
\end{equation}

Finally, we note that simple phenomonelogical derivations can be given
for the sum rules \ref{eq:appone} and \ref{eq:appresult}.  Since $f_0$
is a local density-density interaction, the inclusion of the Coulomb
interaction is in some sense equivalent to taking $f_0 \rightarrow f_0
+ v(q)$.  As we take $q$ to zero, this translates to the effective
divergence of $f_0$.  If we allow $f_0$ to diverge in sum rule Eq.
\ref{eq:appzero} we immediately obtain Eq.  \ref{eq:appone}.

The derivation for Eq. \ref{eq:appresult} proceeds along a similar
line.  The Chern-Simons interaction as written in Eq. \ref{eq:Udef} is
an interaction between transverse current and density.  Using current
conservation $\omega j_0 = q j_x$ we can write this as an interaction
between the transverse and the longitudinal currents (Here $\hatn x$
is chosen to be the longitudinal direction).  We thus rewrite this
interaction energy as
\begin{equation}
  \delta E = j^* U j = \frac{2 \pi \tilde i \phi}{\omega} \left[ j_x^*
  j_y - j_y^* j_x) \right]
\end{equation}
Such an interaction term can be represented in Fermi liquid theory by
letting $ f_1 \rightarrow f_1 + \frac{2 \pi i \tilde \phi}{\omega}$
and $f_{-1} \rightarrow  f_{-1} - \frac{2 \pi i \tilde \phi}{\omega}$.
Note that $f_1$ being complex is a reflection of the fact that the
Chern-Simons interaction is not time reversal invariant.  Letting
$\omega$ and $q$ both go to zero then results in the effective
divergence of $f_0$, $f_1$ and $f_{-1}$, thus yielding Eq.
\ref{eq:appresult} from Eq. \ref{eq:appzero}.

\end{document}